\documentstyle[preprint,pra,aps]{revtex}
\begin{document}
\draft
\title{\bf
\qquad \qquad \qquad \qquad \qquad
[Phys. Lett. B, 1995 to be published]
\\
Limits of Time-Reversal Violating Interaction
from Compound Nuclear Experiments
}
\author{O.~K.~Vorov}
\address{School of Physics, University of New South Wales,
Sydney 2052,
Australia
\\
{\it and}
Nuclear Physics Department,
School of Physics and Astronomy,
%\\
Tel Aviv University,
Ramat Aviv, 69978 Tel Aviv, Israel
\footnote{
Address after October 1995 \\
Fax:  972 (3) 642 9306,
E-mail: ovorov@ccsg.tau.ac.il\\
}}
\date{August 29, 1995}
\maketitle
\begin{abstract}
Mean square matrix elements of the time reversal invariance violating
(TRIV) interaction
between the compound nuclear states
are calculated within the statistical model, using
the explicit form of the TRIV interaction via the $\rho$-meson exchange.
{}From the comparison of the calculated values with
the data known for $p+ ^{27}Al \Leftrightarrow ^{24}Mg + \alpha$ reaction,
and for $\gamma$-correlation measurements in
$^{113}Cd(n,\gamma)^{114}Cd$ process,
the bounds on the TRIV
constant
are obtained ${\bar g}_{\rho} \alt 1.8 \times 10^{-2}$
and ${\bar g}_{\rho} \alt 1.1 \times 10^{-2}$.
The sensitivity of the recently proposed
detailed balance test
experiments on isolated resonances
in $^{32}S$
to the value of ${\bar g}_{\rho}$
is shown to be as high as to reach
values
${\bar g}_{\rho} \sim 10^{-4}$.
\end{abstract}
%\vspace{10mm}

The search for the effects of
direct violation
of the Time Reversal Invariance (TRI)
and the related theoretical studies
still remain to be
a very actual direction of nuclear physics
\cite{HJ,EMW,H,MOLDAUER,SIMONIUS,HHM,EGH,MBBDS,DBMS,CK,EFS,BHW},
\cite{FKPT,SA,KABIR,STODOLSKY,BG,B,WRB,BDGGRS,K,HHR,BSSF}.
Different kinds of experiments
(nuclear, atomic and molecular) are competing in this field
to obtain the tighter bounds on the TRI violating interaction.

It has been realized \cite{SIMONIUS,HHM,EGH}
that the TRIV (P even) nucleon-nucleon
interaction, $V^{TRIV}_{\rho}$,
can be well approximated
by the $\rho$-meson exchange,
at least at the tree level.
(The contributions of the lighter mesons are forbidden
by the Simonius theorem \cite{SIMONIUS}, while
the effects due to heavier mesons should be suppressed by the core
repulsion.)
This interaction can be constructed uniquely \cite{SIMONIUS},
\cite{HHM},\cite{EGH} and it is characterized by the only dimensionless
constant ${\bar g}_{\rho}$ that measures the ratio of the TRI
violating $\rho N N$ amplitude to the normal TRI preserving one,
$g_{\rho}$
\cite{HHM}.
Recently, the parameter ${\bar g}_{\rho}$ has been adopted as
the natural universal measure of the TRI violation in the nuclear
processes,
while the $\rho$-meson-mediated TRIV interaction is used
in the microscopical calculations of the TRIV observables.

Haxton, H\"oring and Musolf \cite{HHM} analyzed the limits
on the value of ${\bar g}_{\rho}$ resulting from the existing
experimental information obtained in atomic and
nuclear experiments.
Engel, Gould and Hnizdo \cite{EGH} used the interaction $V^{TRIV}_{\rho}$
to obtain the effective TRI violating nuclear optical potential.
That enabled them
to calculate microscopically
the ``fivefold'' TRIV correlation
$(\vec{s} \cdot \vec{I} \times \vec{p})(\vec{I} \cdot \vec{p})$
and, by comparison of the result to the experimental value for
the neutron scattering on the aligned $^{165}Ho$ target \cite{K},
to extract the limits on the value of the constant ${\bar g}_{\rho}$.

However, the experimental limits on the TRIV interaction scale
obtained so far, are still several order of magnitude greater
than the values one can expect from the theoretical considerations
\cite{CK},\cite{EGH}.
Recently, the work \cite{EFS} reported on the results of the
TRIV scale obtained in from consideration of the effective Lagrangians.
According to the results of Ref.\cite{EFS}, no TRIV effects could be
expected above the level ${\bar g}_{\rho} \sim 10^{-6}$
(even the stringent estimates were obtained in \cite{CK}),
and this makes
a serious challenge to the contemporary experimental
techniques.

The aim of this work is to show that the limits on the constant
${\bar g}_{\rho}$ that can be extracted from the statistical
nuclear reaction of Ref.\cite{BDGGRS}
and the $\gamma$-ray correlation experiments \cite{BSSF}
are
close, and they can compete with the constraints obtained
from the atomic dipole moment data
(${\bar g}_{\rho} \alt 9.3 \cdot 10^{-3}$)
\cite{HHM},
and to
show that in principle much stricter
bounds on the value of constant ${\bar g}_{\rho} \alt 10^{-4}$
can be reached
in the experiments with compound nucleus $^{32}S$,
which were
proposed by Drake, Mitchell {\it et al.} in
Refs.\cite{MBBDS},\cite{DBMS}.

In the compound nuclear reactions, the symmetry breaking effects
are usually enhanced
\cite{EMW},\cite{SF},\cite{BG,B}, e.g., in the case
of the spatial parity violation
\cite{SF,AL,F,JBY,AB}.
The TRIV observables in compound states are known to be
expressed through the mean square matrix element of
the TRIV interaction
\cite{EMW,FKPT,KABIR,STODOLSKY,BG}.
Using the explicit form
of the TRIV interaction Hamiltonian via the $\rho$-meson exchange
\cite{EGH},\cite{HHM},\cite{EFS},
we make the microscopic calculation of the means square matrix element
(MSME) of the $V^{TRIV}_{\rho}$ between the compound states
using the statistical method developed in \cite{we},
and express MSME through the constant
${\bar g}_{\rho}$.
The conclusions of the work are drawn from the comparison
of the results obtained for particular compound nuclei
with the data.

We use the TRIV nucleon interaction via the $\rho$-meson exchange
(see Refs. \cite{EGH},\cite{HHM}), that reads
\begin{equation}
\label{HTRIV}
V_{TRIV}={\bar g}_{\rho}
\frac{
g^2_{\rho}\mu_v}{8 \pi m^2}
\{ \vec{\tau}_{1} \times \vec{\tau}_{2} \}_z
\frac{e^{-m_{\rho}|r_1-r_2|}}{|r_1-r_2|^3}
(1+m_{\rho}|r_1-r_2|)
\nonumber\\
(\vec{\sigma}_1-\vec{\sigma}_2)\times (\vec{r}_1-\vec{r}_2)
\cdot (\vec{p}_1-\vec{p}_2)
\end{equation}
where ${\bar g}_{\rho}$ is the dimensionless strength parameter
that measures TRIV $\rho N N$ coupling in terms of ordinary
TRI conserving coupling constant $g_{\rho}=2.79$,
$m_{\rho}$ and $m$ are the $\rho$-meson and the nucleon masses
respectively, $\mu=3.7$ is the isovector nucleon magnetic moment
in units of the nuclear magneton, $\vec{r}_{1,2}$ and $\vec{p}_{1,2}$
stand for the coordinates and momenta of the nucleons $1$ and $2$.
$\vec{\sigma}_{1,2}$ and $\vec{\tau}_{1,2}$ are the spin and isospin
Pauli matrices.
The sign $\times$ means the external vector product.

Due to the factor $\{ \vec{\tau}_{1} \times \vec{\tau}_{2} \}_z$,
only the exchange terms of the interaction (\ref{HTRIV}) are
operative. It is therefore convenient to rewrite
the interaction in the form $V'_{TRIV}$, so that only the direct terms
should be taken into account
\begin{equation}
\label{SECOND}
V^{\rho}_{TRIV} \quad = \quad \frac{1}{2}\sum_{ab,cd}
a^{+}bV'^{\quad ab,cd}_{TRIV}c^{+}d,
\end{equation}
Using the Fiertz transformation for the
spin and isospin matrices,
\begin{displaymath}
\{ \vec{\tau}_{1} \times \vec{\tau}_{2} \}_3 \rightarrow
i(\vec{\tau}_{1} - \vec{\tau}_{2} )_3, \quad
(\vec{\sigma}_{1} - \vec{\sigma}_{2})_3 \rightarrow
-i \{ \vec{\sigma}_{1} \times \vec{\sigma}_{2} \}_3
\end{displaymath}
the matrix elements $V'^{ab,cd}_{TRIV}$ can be found in the form
\widetext
\begin{eqnarray}
\label{HTRIVdir}
V'^{ab,cd}_{TRIV}=i{\bar g}_{\rho}
\frac{g^2_{\rho}\mu_v}{2 m^2 m^2_{\rho}}
\int d 1 d 2 \cdot
\qquad \qquad \qquad \qquad \qquad \qquad
\nonumber\\ \cdot
\psi^{\dagger}_a(\vec{r}_1) \psi^{\dagger}_c(\vec{r}_2)
(\vec{\tau}^{ab}_z - \vec{\tau}^{cd}_z)
(\vec{\nabla}_1-\vec{\nabla}_2)_{\alpha}
\left[ \vec{\nabla}_{1 \beta},
\frac{m_{\rho}^2 e^{-m_{\rho}|r_1-r_2|}}{4 \pi |r_1-r_2|}
\right]
(\vec{\sigma}^{ab}_{\alpha}\vec{\sigma}^{cd}_{\beta} -
\vec{\sigma}^{ab}_{\beta}\vec{\sigma}^{cd}_{\alpha})
\psi_d (\vec{r}_1) \psi_b (\vec{r}_2),
\end{eqnarray}
\narrowtext
where $\vec{\nabla}_1=i \vec{p}_1$,
$d1$ and $d2$ mean integration and summation over the spatial
and the spin variables,
$[ \quad , \quad ]$ means the
commutator and the summation over the
double indices is assumed.
Like the ordinary (symmetry preserving)
strong interaction,
the {\it proton-neutron} interaction (\ref{HTRIV}),(\ref{HTRIVdir})
has P-even
selection rules,
with the essential difference that the matrix
elements of $V_{TRIV}$ are purely imaginary,
and therefore, its diagonal matrix elements vanish.

{\bf 1.} The method of calculation of mean squared weak matrix element
is based on the equivalence theorem of canonical and microcanonical
ensembles for a system with a large number of degrees of freedom.
It has
been developed previously for the microscopic calculation of the
parity violating matrix elements between the compound states
\cite{we}. (Later, it has been applied to the calculation of the
P and T noninvariant matrix elements \cite{PTP}).
We recall
that the wave function of any compound state with angular
momentum $J$ and parity $\pi$, $|J^{\pi})$,
can be expressed
as the superposition of the simple components $|J^{\pi}>$ \cite{BM}
\begin{equation}
\label{COMPOUND}
|J^{\pi})=\sum_{\alpha} C_{\alpha}|J^{\pi}_{\alpha}>, \qquad
|J^{\pi}_{\alpha}>=(a^+bc^+de^+ ...)_{J^{\pi}}|G.s.>
\end{equation}
Here,
$|J^{\pi}_{\alpha}>$
are the multi-particle excitations over
the shell-model
ground state $|G.s.>$.
By $(...)_{J^{\pi}}$, we
mean the coupling of nucleon creators $a^{+}$ and destructors $a$
to total angular momentum $J$ and
parity $\pi$.
The energies $E_{\alpha}$ of the ``principal'' components
$|J^{\pi}_{\alpha}>$
which dominates the
normalization of (\ref{COMPOUND}),
must be within the interval $[{\cal E}-\frac{\Gamma
_{spr}}{2},{\cal E}+\frac{\Gamma_{spr}}{2}]$ \cite{BM},\cite{SF},
where ${\cal E}$ is the energy of the compound
state and $\Gamma_{spr}$ is the spreading width
(typically, $\Gamma_{spr} \sim 1..3 MeV$, Refs.\cite{BM}).
These components
(which contain several excited nucleons) can be composed by excitations
of protons and neutrons only inside
incomplete valence shells.
Mean squared values of the coefficients $C(E_{\alpha})$
can be described by the formulae (see e.g., Ref.\cite{BM})
\begin{eqnarray}
\label{MICRO}
\overline{C^{2}(E_{\alpha})}=\frac{1}{{\cal N}}\Delta(\Gamma_{spr},
{\cal E}-E_{\alpha}),\qquad {\cal N}=\frac{\pi\Gamma_{spr}}{2d},\\
\nonumber
\Delta(\Gamma_{spr},{\cal E}-E_{\alpha})=
\frac{\Gamma_{spr}^{2}/4}{({\cal E}-E_{\alpha})^{2}+\Gamma_{spr}^{2}/4},
\end{eqnarray}
where $E_{\alpha}$ is the energy of an arbitrary many-particle
configuration),
$d$ is the averaged energy distance between the resonances,
and ${\cal N} \gg$$ 1$
is the number of principal components.
The Breit-Wigner-type factor $\Delta$ describes quenching of weights
for states distant in energy.
It is a ``spread''
$\delta$-function normalized as to be of
order unity for $|{\cal E}-E_{\alpha}|\alt\Gamma_{spr}/2$, and
obeying the
limit
$\Delta(\Gamma_{spr},{\cal E}-E_{\alpha})
\to$ $\frac{\pi\Gamma_{spr}}{2}\delta({\cal E}-E_{\alpha})$
for $\Gamma_{spr}\to 0$.
Thus
the average with weight $\overline{C^2}$ is equivalent to
the microcanonical average.

The TRIV interaction (Eq.(\ref{HTRIV})) is a two-body operator
(\ref{SECOND}), and it has nonzero matrix elements between the
single-particle states of the valence shells.
Its matrix elements between the compound states
(\ref{COMPOUND})
can be expressed in terms
of the single-particle matrix elements $V'^{ab,cd}_{TRIV}$
by using of the statistical properties of the components of
the compound states
\footnote{
The case of the parity-odd interaction
considered in \cite{we} is a bit more involved.
The most important contribution to the parity odd MSME is given by
the single-particle symmetry breaking potential
which acts in combination
with the symmetry preserving residual strong interaction to dominate the MSME.
In the present T-odd case, the corresponding single-particle
potential that can be obtained from
Eqs.(\ref{HTRIV},\ref{SECOND},\ref{HTRIVdir}),
does not contain coherent contribution from the core nucleons
\cite{EGH}. Therefore, the T-odd MSME is dominated by the
matrix elements of (\ref{HTRIV}),(\ref{HTRIVdir}).
}.
To calculate the mean squared value of the TRIV matrix element
$\overline{|V^{2}_{TRIV}|} \equiv v^2_{TRIV}$ one can write it in the form
\begin{eqnarray}
\label{DOUBLE}
v_{TRIV} \quad = \quad
\overline{|(J^{\pi}|V^{\rho}_{TRIV}|{J^{\pi}}')
({J^{\pi}}'|V^{\rho}_{TRIV}|J^{\pi})|} \quad =
\nonumber\\=
\quad \sum_{\alpha \beta} \overline{|C_{\alpha}' C_{\beta}'
(J^{\pi}|V^{\rho}_{TRIV}|J^{\pi}_{\alpha}>
<J^{\pi}_{\beta}| V^{\rho}_{TRIV}|J^{\pi})|} \quad
\end{eqnarray}
using the expansion of the compound state $|J'^{\pi})$
in terms
of their multiparticle components (\ref{COMPOUND}).
Here $|J'^{\pi})$ and $|J^{\pi})$ are the different compound states
with the same quantum
numbers.
The coefficients $C_{\alpha}$ are statistically independent
\cite{BM},
$\overline{C_{\alpha}C_{\beta}} = \overline{C^{2}_{\alpha}}
\delta_{\alpha \beta} =
\delta_{\alpha \beta} \frac{1}{{\cal N}}
\Delta(\Gamma_{spr},{\cal E}-E_{\alpha})$, [Eq.(\ref{MICRO})].
Therefore, in (\ref{DOUBLE}) we obtain
\begin{eqnarray}
\label{DOUBLE1}
v_{TRIV} \quad = \quad
\sum_{\alpha}
\frac{1}{{\cal N}}
\Delta(\Gamma_{spr},{\cal E}-E_{\alpha})
\times
\nonumber\\ \times
\overline{|(J^{\pi}|V^{\rho}_{TRIV}|J^{\pi}_{\alpha}>
<J^{\pi}_{\alpha}|V^{\rho}_{TRIV}|J^{\pi})|},
\end{eqnarray}
In the second quantization representation (\ref{SECOND}),
summation over the intermediate states $\alpha$
in the last equation
is equivalent to summation over different components of the operator
$V^{\rho}_{TRIV}$.
In the expression for the remaining average,
$\overline{ (J^{\pi}|V^{\rho}_{TRIV}V^{\rho}_{TRIV} |J^{\pi})}$,
in (\ref{DOUBLE1}), the weights $\overline{CC}$
will appear in the same way as above.
To calculate this average over the resonances $|J^{\pi})$,
we now use, instead
of the microcanonical ensemble (\ref{MICRO}),
the equivalent canonical (Gibbsean) ensemble.
(As we know from the statistical mechanics, the latter can be used as
an equivalent counterpart of the microcanonical ensemble,
provided ${\cal N} \gg 1$.)
Thereby, the average expectation value
in (\ref{DOUBLE1}) can be evaluated
using
the standard contractor rules for the Fermi operators,
given by Wick theorem
$\overline{(J^{\pi}|a^{+}b|J^{\pi})}=
(\overline{ a^{+}b })_{T}
=\delta_{ab}\nu_{a}(T)$, for
$\nu_{a}(T)$ being the finite temperature Fermi occupation probabilities
of the single-particle levels,
$\nu_{a}(T)=\lbrace exp[(\varepsilon_{a}-\lambda)/T]+1
\rbrace ^{-1}$.
The parameters of the effective canonical ensemble,
$T$ (the temperature), and $\lambda_{p}$, and $\lambda_{n}$
(proton and neutron chemical potentials)
are to be determined from the ``consistency
`` equations $E = \sum_{a}\nu_{a}\varepsilon_
{a}$, $ Z=\sum_{p}\nu_{p}$, and $N=\sum_{n}\nu_{n}$
for the total energy $E$,
nuclear charge $Z$, and neutron number $N$ correspondingly.
Making use of the Wick theorem reduces (\ref{DOUBLE}) to the
result (see \cite{we}):
\begin{eqnarray}
\label{RESULTa}
v_{TRIV} \quad = \quad \sqrt{\overline{|V^{2}_{TRIV}|}}=
\qquad  \qquad \qquad \qquad \qquad \qquad \\ \nonumber
\sqrt{\frac{d}{\pi\Gamma_{spr}}} \left\{ \sum_{abcd}\nu_{a}(T)[1-\nu_{b}(T)]
\nu_{c}(T)[1-\nu_{d}(T)]
\mid  V'^{\quad ab,cd}_{TRIV} \mid ^{2}\Delta(\Gamma_{spr},
\varepsilon_{a}-\varepsilon_{b}+
\varepsilon_{c}-\varepsilon_{d})\right\} ^{\frac{1}{2}}.
\end{eqnarray}
Here, $V'^{\quad ab,cd}_{TRIV}$ is the direct matrix element
of the TRIV Hamiltonian in the form (\ref{HTRIVdir}), defined according to
Eq.(\ref{SECOND}).
The factor $\Delta$ reflects the required energy balance
in the transitions due to $V'_{TRIV}$ and
ensures that the components of both
the states $|J^{\pi})$ and $|J'^{\pi})$ are in the same energy domain
${\cal E} \pm \Gamma_{spr}/2$.

The numerical calculations
of the $v_{TRIV}$
have been performed with the use
of single-particle basis of states obtained by numerical calculations
of the eigenvalue
problem for the Woods-Saxon potential with spin-orbital
interaction  in the
form
$U(r)=-U_{0}f(r)+
U_{ls}(\vec{\sigma} \vec{l})(\hbar/(m_{\pi}c))^{2}\frac{1}{r}
\frac{df}{dr}+ U_{c}$ with
$f(r)=(1+e^{(r-R)/a})^{-1}$,
where
$\vec{l}$ is the orbital angular momentum, $U_{c}$ means
Coulomb correction for protons, $U_{c}=3Ze^{2}/(2R)(1-r^{2}/(3R^2)),
r\leq R$ and $U_{c}=Ze^{2}/r, r>R$, for $R$, $a$, and $r$
being the nuclear radius,
diffusity parameter and radial variable correspondingly.
The parameter values
were used in accordance with Bohr-Mottelson prescription
\cite{BM}.
As the interaction via the exchange by the massive $\rho$-meson
is affected by the
repulsion at the core scale distances,
the short-range correlations were accounted for
within the standard Jastrow method (see, e.g., \cite{SAD}).

{\bf 2}. There were two experiments with reaction
$p + ^{24}Mg \Leftrightarrow ^{27}Al + \alpha$
to study TRIV via detailed balance test
\cite{WRB},\cite{BDGGRS}.
They give upper limit on the value
of the parameter $\alpha$
that is
the ratio of the r.m.s.
of the TRIV matrix elements between the compound states
to the regular TRI-preserving strong interaction $V_S$:
\begin{equation}
\label{ALPHA}
\alpha = \frac{v_{TRIV}}{\left[ \overline {V_S^2} \right]^{1/2}}.
\end{equation}
The revision of the data of the improved experiment \cite{BDGGRS}
by Harney, H\"upper and Richter \cite{HHR}
gave the following limit:
\begin{equation}
\label{LIMIT}
\alpha \quad \alt \quad 2.6 \cdot 10^{-4}
\qquad (80 \quad \% \quad conf.lev.).
\end{equation}
To obtain the theoretical value of the quantity $\alpha$
in terms of the universal constant
${\bar g}_{\rho}$, we need also the r.m.s. of
the residual strong interaction
matrix elements between the multiparticle
components of the compound states.
We have employed the most widely used Landau-Migdal
particle-hole
interaction
which rises
to Landau Fermi liquid theory.
For the case of a nucleus it
was justified in the Theory of Finite Fermi Systems
\cite{Migdal,NUC}.
In explicit form, the Landau-Migdal interaction reads
\begin{equation}
\label{V0}
V_S
=C\delta(\vec{r}_1-\vec{r}_2)[f+f'(\vec{\tau}_1 \vec{\tau}_2)+
g(\vec{\sigma}_1 \vec{\sigma}_2)+
g'(\vec{\tau}_1 \vec{\tau}_2)(\vec{\sigma}_1 \vec{\sigma}_2)],
\end{equation}
where $C=300$ $MeV$ $fm^{3}$ is the universal
Migdal constant \cite{Migdal},\cite{NUC},
$f,f',g,g'$ are the density dependent
strengths
$f=f_{in}-(f_{ex}-f_{in})(\rho(r)-\rho(0))/\rho(0)$,
$\rho(r)=\rho(0)f(r)$.
The interaction (\ref{V0}) has been successfully used by many authors
(see Refs. \cite{NUC}) to describe a great amount of
various properties of nuclei at quantitative level.
The values of the parameters
[$f_{ex}=-1.95$, $f_{in}=-0.075$ $f'_{ex}=0.05$ $f'_{in}=0.675$,
$g_{in}=g_{ex}=0.575$,
and $g'_{in}=g'_{ex}=0.725$]
were chosen to fit a variety of
experimental data.
It is therefore reasonable to believe that the parametrization
(\ref{V0})
describes strength and the main properties of the
residual interaction correctly.

The value of $\left[\overline{V_S^2}\right]^{1/2}$ can be calculated
via the same formula as $v_{TRIV}$ (\ref{RESULTa}) with the substitution
the direct matrix elements of (\ref{V0}) into (\ref{RESULTa})
instead of $V'^{\quad ab,cd}_{TRIV}$, (\ref{HTRIVdir}).
In this case, we need in fact to calculate the MSME of $V$ between the
components of the compound states $\{ |J^{\pi}> \}$ (\ref{COMPOUND})
rather than between the
compound states $\{ | J^{\pi}) \}$. However, we can use the same
formula (\ref{RESULTa}), as the sum of the squared matrix elements
taken in the energy interval $\Gamma_{spr}$
is independent of reference basis,
while
the same thermodynamical
expression for the occupation numbers $\nu$ can be used \cite{BM}
in averaging over the unmixed components.
In contrast to the TRIV interaction, the normal strong
interaction that is given by the direct terms of (\ref{V0})
contain the the diagonal matrix elements
$\langle J^{\pi}|V_{S}|J^{\pi}\rangle$.
Those have to be excluded from the sum (\ref{RESULTa}).

In the expression for $\alpha$ (\ref{ALPHA}),
all the coefficients in front of the square root
of the sum in (\ref{RESULTa})
are canceled.
Using the value of temperature $T=2.3 MeV$ to reach the excitation
energy $E_{exc} \sim 16..20 MeV$ of the compound nucleus $^{28}Si$
\cite{BDGGRS} ($\Gamma_{spr} \simeq 1$MeV \cite{BDGGRS}),
we have obtained the value of $\alpha$, and, consequently,
the upper limit on the TRIV constant ${\bar g}_{\rho}$, as
follows
\begin{equation}
\label{LIMITc}
\alpha =  1.44 \cdot 10^{-2} {\bar g}_{\rho} ,
\qquad \qquad {\bar g}_{\rho} \quad \alt
\quad 1.8 \cdot 10^{-2}.
\end{equation}
This limit is close to the one
obtained in Ref.\cite{HHM} from the atomic dipole moment data
(${\bar g}_{\rho}  \alt 0.93 \cdot 10^{-2}$).
It should be borne in mind that
twelve years passed after the experiment of Ref.\cite{BDGGRS}.

{\bf 3}. We consider now the result of the
novel experiments \cite{BSSF} where the
forward-backward asymmetry in individual $\gamma$-ray transitions
resulting from unpolarized neutron capture in $p$-resonances
by the target $^{113}Cd$
has been measured.
(The possibility of measurement of T-odd correlations
in the experiments
of such type was
discussed earlier in Ref.\cite{SF85}.)
The value of TRIV mixing coefficient $\beta$ has been extracted
for the compound nucleus $^{114}Cd$
(neutron $p$-resonance at energy ${\cal E}_{p1} = 7.04 eV$)
to be
\begin{equation}
\beta
\simeq
\left| \frac{v_{TRIV}}{{\cal E}_{p1}-{\cal E}_{p2}} \right|,
\qquad \beta^{EXP} \alt 0.08
\end{equation}
Assuming the mixing with the closest $p$-resonance
(${\cal E}_{p1}=21.8$ eV \cite{FSPHHW}),
we obtain,
for
the value of TRIV matrix element $v_{TRIV} \alt 1.18 eV$.
The present calculation of the root MSME $v_{TRIV}$ for the
low energy neutron resonances in $^{114}Cd$
(the temperature $T=1.04 MeV$, $d=27 eV$, and $\Gamma_{spr}=3MeV$)
gives
the following quantity and corresponding bound
\begin{equation}
\label{CADMIUM}
v_{TRIV} = {\bar g}_{\rho} \cdot 0.11 \cdot 10^{3} eV,
\qquad
{\bar g}_{\rho} \alt 1.1 \cdot 10^{-2},
\end{equation}
This constraint is tighter than in the case  of $^{28}Si$
(\ref{LIMITc}).
One should note that the bound on ${\bar g}_{\rho}$
extracted from $\beta$,
depends strongly on the density of the $p$-resonances.
E.g., suppose the same upper limit on $\beta$ (\ref{CADMIUM})
to be obtained on
one of the resonances at
${\cal E}_{p} = 98.5$ eV and ${\cal E}_{p} =102.3$ eV \cite{FSPHHW}.
This would result in the lower bound on the TRIV constant:
${\bar g}_{\rho} \alt (2-3) \times 10^{-3}$.

To judge on the reliability of the statistical method used here,
it is interesting to see the results the method gives for the
P-odd, TRI preserving weak interaction for
the compound systems similar to those considered here, where
the experimental data are known.
(Earlier, the method
produced the result consistent with experimental data for $^{233}Th$
compound nucleus \cite{we}).
For $^{28}Si$ and $^{32}S$ (discussed below),
the ``closest case'' is the compound nucleus $^{36}Cl$, where
the experimental value of
the P violating MSME  $v_{PV}=60(\pm 20)meV$ \cite{ANTONOV} was
obtained in the low energy resonances.
The theoretical result obtained by the present method
(as described in \cite{we}), is $26 meV$.
For cadmium, there are data in low energies for the same isotope,
$^{114}Cd$ \cite{AL}: $v_{PV}=1.4 (\pm 0.5)meV$.
The corresponding theoretical value is found to be $1.2 meV$,
for the same values of $d,\Gamma_{spr}$ and $T$ as above.
Therefore, the method seems to
give reliable results.

{\bf 4}. Now, let us estimate the scale of the constant ${\bar g}_{\rho}$
which can be available for investigation in the experiments with
with formation of the compound nucleus
$^{32}S$ in the energy region of separated resonances, proposed
in \cite{MBBDS},\cite{DBMS} (detailed balance test in reaction
$p+^{31}P \Leftrightarrow ^{28}Si + \alpha$).
In this case, the observable T-odd quantity is the asymmetry
parameter $\Delta$ that measures
the degree of the detailed balance violation \cite{MBBDS},
$\Delta = 2 \quad
\frac{ \sigma_{p,\alpha} - \sigma_{\alpha,p}}
{\sigma_{p,\alpha} + \sigma_{\alpha,p}}$.
Here, $\sigma_{p,\alpha}$ and $\sigma_{\alpha,p}$
are the properly weighted differential cross sections
for direct and inverse processes, see \cite{MBBDS}.
As was shown in Refs.\cite{MBBDS},\cite{DBMS},
the quantity $\Delta$
is proportional to the average value of the matrix element of the
T-odd interaction $v_{TRIV}$
\begin{displaymath}
\Delta = {\cal F} \cdot v_{TRIV},
\end{displaymath}
where the factor ${\cal F}$ depends in a complicated
way on the angles and resonance parameters.
Its absolute value $|{\cal F}|=|\frac{\Delta}{v_{TRIV}}|$
determines the sensitivity of the experiments to the T-odd
interaction.
The analyis of Refs.\cite{MBBDS},\cite{DBMS}
has shown that high sensitivity can be produced by an appropriate
choice of the conditions of measurement (energies, angles {\it etc}).

Basing on the the present results for the $v_{TRIV}$
[Eq.(\ref{RESULTa})]
and on
the values of $|{\cal F}|$ obtained in Refs.\cite{MBBDS},\cite{DBMS},
we can make the
estimation
for the upper bound on the TRIV constant ${\bar g}_{\rho}$
that could be reached in such experiments.
The values of the resonance energy spacing and the spreading width
were take from Ref.\cite{DBMS}:
$d = 7.2 \cdot 10^4$ eV and $\Gamma_{spr} = 1 MeV$.
The value of temperature $T=2.1 MeV$ was used
in accordance to the consistency
condition for excitation energy (see above) to reach the excitation
energy $E_{exc} \sim 10..12 MeV$ of the compound nucleus \cite{DBMS}.
For the `best sensitivity' in range
$|\frac{\Delta}{v_{TRIV}}| \sim (1.5-15) \cdot 10^{-3}$
\cite{MBBDS},\cite{DBMS},
the results are summarized as
follows:

\newpage

\parbox{40mm}{
Level of the \\
measurable effect\cite{DBMS}
}
\parbox{40mm}{
$v_{TRIV}$ expected
}
\parbox{40mm}{
Present result \\
for the $v_{TRIV}$
}
\parbox{35mm}{
Bound on ${\bar g}_{\rho}$ \\
as resulted
}

\vspace{5mm}
\parbox{37mm}{
$\Delta \sim 10^{-2}$
}
\parbox{37mm}{
$6.7 - 0.67$ eV
}
\parbox{37mm}{
${\bar g}_{\rho}4.8 \cdot 10^3$ eV
}
\parbox{37mm}{
${\bar g}_{\rho} \alt 1.4\cdot (10^{-3}-10^{-4})$
}

\vspace{5mm}

Therefore, it can be concluded that the experiments for
detail balance tests of TRI with $^{32}S$-compound nucleus
proposed in Ref.\cite{DBMS} can produce not only the
new bound on the TRIV constant ${\bar g}_{\rho}$, but
lower the upper limit for ${\bar g}_{\rho}$
down to
$10^{-3}-10^{-4}$.
This is already close to the most ``pessimistic''
expectations for the value of constant ${\bar g}_{\rho}$ that
follow from the considerations by Engel and collaborators \cite{EFS}.

In summary, the mean square matrix elements
of the TRIV interacting between the highly excited nuclear states
have been calculated
within the statistical approach, starting with the microscopic
TRIV nucleon interaction via $\rho$-meson exchange.
The r.m.s. values of the TRIV matrix elements are expressed
through the $\rho$-mediated interaction constant.
Particular calculations have been made for the
$^{28}Si$, $^{114}Cd$ and $^{32}S$ compound nuclei.
The results, being compared with the experimental data
from the compound nuclear reactions, allow one to obtain
the limits on the TRVI constant
${\bar g}_{\rho} \alt 1.8 \times 10^{-2}$
and ${\bar g}_{\rho} \alt 1.1 \times 10^{-2}$,
for the detailed balance test
experiment in the statistical reaction in
the region of overlapping resonances
\cite{BDGGRS} and in the
$\gamma$-correlations in low-energy neutron resonances \cite{BSSF},
respectively.
The sensitivity of the experiments of the detailed balance
tests in the $p+^{31}P \Leftrightarrow ^{28}Si + \alpha$
reaction, proposed
in \cite{MBBDS},\cite{DBMS} to the value of TRIV $\rho$-meson
constant is found high enough to study the new range
of values ${\bar g}_{\rho} \sim 10^{-4}$.

The conclusion can be drawn that the nuclear experiments
with compound states seem rather prospective and
are able to provide important information on the TRIV
scale.
Of course, the TRIV matrix element between the compound states
is a fluctuating quantity, and hence the bounds obtained from
$v_{TRIV}$ are approximative.
A more definite information can be obtained from the measurements
on several resonances and/or in different compound systems.

One of the interesting issues related to the TRIV interaction
via the $\rho$-meson exchange (\ref{HTRIV}) is that
explicit microscopic form of the TRIV Hamiltonian opens the possibility
to study correlations between TRIV matrix elements and the
matrix elements of the normal TRI preserving strong interaction.
(Usually, the TRI preserving and TRI violating matrix elements
have been treated as statistically independent
random quantities, i.e., the members of the Gaussian unitary ensemble.)
Apparently, these correlations should be depressed for the same
reason as the correlation between P-odd and P,T-odd matrix
elements \cite{PTP}.
However, this interesting question deserves studies.

The author is grateful to V.V.Flambaum for his interest to the work
and comments,
V.F.~Dmitriev and V.B.~Telitsin for kindly providing
the author with the codes to solve eigenvalue problem
in Saxon-Woods potential,
and to V.A.Kuznetsov for reading the manuscript.
The work has been supported by ARC grant.

\noindent
\end{document}